\documentclass[english,prd,onecolumn,superscriptaddress,nofootinbib,preprintnumbers]{revtex4}
\usepackage[latin1]{inputenc}
\usepackage{graphicx}
\usepackage{amssymb}

\usepackage{amsfonts}
\usepackage{dcolumn}
\usepackage{hyperref}

\def\be{\begin{equation}}
\def\ee{\end{equation}}
\def\beq{\begin{eqnarray}}
\def\eeq{\end{eqnarray}}

\newcommand{\rd}{{\rm d}}

\makeatother

\usepackage{babel}
\makeatother
\begin{document}

\title{Galileon gravity and its relevance to late time cosmic acceleration}

\author{Radouane Gannouji}
\affiliation{IUCAA, Post Bag 4, Ganeshkhind, Pune 411 007, India}

\author{M. Sami}
\affiliation{Centre of Theoretical Physics, Jamia Millia Islamia, New Delhi-110025, India}

\begin{abstract}
We  consider the covariant galileon gravity taking into account the
third order and fourth order scalar field Lagrangians $L_3(\pi)$ and
$L_4(\pi)$ consisting of three and four $\pi$'s with four and five
derivatives acting on them respectively. The background dynamical
equations are set up for the system under consideration and  the
stability of the self accelerating solution is demonstrated in
general setting. We extended this study to the general case of the
fifth order theory. For spherically symmetric static background, we
spell out conditions for suppression of fifth force effects mediated
by the galileon field $\pi$. We study field perturbations in the
fixed background and investigate conditions for their causal
propagation. We also briefly discuss metric fluctuations and derive
evolution equation for matter perturbations in galileon gravity.
\end{abstract}
%


\maketitle

\section{Introduction}
The phenomenon of late time cosmic
acceleration \cite{Perlmutter:1998np,Riess:1998cb,Seljak:2004xh,Spergel:2006hy}
is as challenging theoretically as was the problem of black body
radiation whose resolution unveiled many secrets of micro physics.
 At present, there is no definite clue
for the theoretical understanding of the nature of cosmic repulsion.
In recent years, a variety of approaches have been employed to
attack the problem.
 According to the standard lore, the late time acceleration can be
accounted for by supplementing the energy momentum tensor by an
exotic fluid component with large negative pressure dubbed {\it dark
energy} \cite{rev1,rev}
 The simplest candidate
of dark energy is provided by cosmological constant $\Lambda$.
However, its small numerical value leads to {\it fine tuning}
problem and we do not understand why it becomes important today {\it
a la} {\it coincidence} problem.

Scalar fields provide an interesting alternative to cosmological
constant though they do not address the cosmological constant
problem. To this effect, cosmological dynamics of a variety of
scalar fields has been investigated in the literature(see
review \cite{rev1} for details). They can mimic cosmological constant
like behavior at late times and can provide a viable cosmological
dynamics at early epochs. Scalar field models with generic features
are capable of alleviating the fine tuning and coincidence problems.
As for the observation, at present, it is absolutely consistent with
$\Lambda$ but at the same time, a large number of scalar field
models are also permitted. Future data should allow to narrow down
the class of permissible models of dark energy.

It is quite possible that there is no dark energy and the late
cosmic acceleration is an artifact of infrared modification of
gravity. We know that gravity is modified at short distance and
there is no guarantee that it would not suffer any correction at
large scales where it is never verified directly. Large scale
modifications might arise from extra dimensional effects or can be
inspired by fundamental theories. They can also be motivated by
phenomenological considerations such as $f(R)$ theories of
gravity \cite{FRB} or the massive theories of gravity. However, any
large scale modification of gravity should reconcile with local
physics constraints and should have potential of being distinguished
from cosmological constant.

 The infrared modified theories of
gravity essentially contain additional degrees of freedom. The f(R)
theories contain a scalar field which mediates fifth force and might
contradicts the local gravity constraints such as the solar system
or laboratory tests. Broadly, two mechanisms for hiding the scalar
field effects locally have been employed in the literature. In
$f(R)$ theories of gravity, the scalar field  is screened via the
so-called chameleon mechanism \cite{Khoury:2003aq}, by making scalar
field mass dependent on the local matter density. In generic models
of f(R) gravity \cite{Hu}, the chameleon mechanism allows to satisfy
the local gravity constraints but at the same time make these models
vulnerable to curvature singularity whose resolution requires the
fine tuning worse than the one encountered in $\Lambda CDM$ model.
The problem can be alleviated by invoking $R^2$ correction but the
scenario becomes problematic if extended to early universe
\cite{Hu}.

An alternative possibility of large scale modification of gravity is
provided by an effective scalar field $\pi$ dubbed galileon \cite{Nicolis:2008in}. In
particular such a field appears in the decoupling limit of DGP. The
Lagrangian of the field respects the so called shift symmetry in a Minkowskian background:
$\pi\to \pi+c$ and $\partial_{\mu}\pi\to \partial_{\mu}\pi+b_{\mu}$
where $c$ and $b_{\mu}$ are constants. Thank to this symmetry, the
equations of motion for the field contain only second derivatives.
In four space time dimensions, there exist five Lagrangians ${L_i},
i=1,5$ where $L_1$ is linear in $\pi$, $L_2$ contains normal kinetic
term. ${L_3}$ involves three $\pi$'s and four derivatives acting on
them. This Lagrangian is obtained in the decoupling limit of DGP.
The fourth and the fifth order Lagrangians involve four $\pi$'s and
six derivatives, five $\pi$'s and seven derivatives acting on the
field respectively. A general covariant form of galileon
Lagrangian is obtained in Ref.\cite{Deffayet:2009wt}(see also Ref.\cite{Chow:2009fm} on the related theme).

 In  DGP or its 4 dimensional generalizations-galileon gravity, the effects of extra degree are suppressed using the  Vainshtein
\cite{Vainshtein:1972sx} mechanism which allows us to recover general
relativity at small scales due to non-linear interaction. From this
point of view, the DGP model is  attractive model which has a
self-accelerating solution,
 an asymptotically de Sitter solution even in the absence of vacuum
energy. Unfortunately this solution suffers from instabilities
\cite{Luty:2003vm,Nicolis:2004qq,Gorbunov:2005zk,Charmousis:2006pn,Dvali:2006if,Gregory:2007xy}.

galileon gravity can give rise to late time acceleration and is
interesting for the following reasons: (i) It is free from negative
energy instabilities. (ii) Unlike  $f(R)$ theories, galileon
modified gravity does not suffer from curvature singularity.
(iii)The chameleon mechanism in $f(R)$ might come into conflict with
the equivalence principle if the test bodies are considered as
extended
 whereas the Vainshtein mechanism is free from this problem \cite{Hui}.

 In this paper we study 4th order galileon gravity including
 $L_ 3$ and $L_4$ terms in the Lagrangian. We set up FRW background dynamics
 and examine the self accelerating solution. We carry out detailed
 investigations on the stability of the solutions and discuss
 the spherical symmetric solutions to check the local suppression
 of $\pi$ effects. We also investigate matter perturbations
 in the model under consideration.

\section{Lowest order Galileon gravity and its self accelerating FRW background}

Recently, an interesting generalization of the DGP action
 in 4D was proposed in Ref.\cite{Nicolis:2008in}.
 The authors considered a consistent general action
with a self interacting scalar field ($\pi$) coupled. It is
remarkable that the action can be motivated by higher dimensional
considerations \cite{Claudia}. In what follows we shall consider the
action is invariant under  Galilean transformation

\be \pi(x)\rightarrow \pi (x)+b_\mu x^\mu +c \ee

For the sake of simplicity, we first examine the galileon model in
the lowest non-trivial order keeping up to third order term $L_3$ in
the Lagrangian,

\be \label{eq:3order} \mathcal{S}=\int{\rd^4
x}\sqrt{-g}\left(\frac{R}{2}+c_1\pi-\frac{c_2}{2}(\nabla\pi)^2-\frac{c_3}{2}
(\nabla\pi)^2
\Box\pi\right)+\mathcal{S}_m[\psi_m,e^{2\beta\pi}g_{\mu\nu}]
\label{action1}
 \ee
Similar expression occurs in the DGP model. The corresponding
Einstein's equations are

\beq
G_{\mu\nu}&=& T_{\mu\nu}^{(m)}+c_1\pi g_{\mu\nu}+c_2\left(\pi_{;\mu}\pi_{;\nu}-
\frac{1}{2}g_{\mu\nu}(\nabla\pi)^2\right)+c_3\left(\pi_{,\mu}\pi_{;\nu}\Box\pi+g_{\mu\nu}\pi_{;\lambda}\pi^{;\lambda\rho}\pi_{;\rho}-\pi^{;\rho}\left[\pi_{;\mu}\pi_{;\nu\rho}+\pi_{;\nu}\pi_{;\mu\rho}\right]\right)\\
0&=&\beta
T^{(m)}+c_1+c_2\Box\pi+c_3\left((\Box\pi)^2-\pi_{;\mu\nu}\pi^{;\mu\nu}-R^{\mu\nu}\pi_{;\mu}\pi_{;\nu}\right)
\label{E1}
 \eeq

where $T^{(m)}$ is the trace of the matter energy-momentum tensor,
 $T_{\mu\nu}^{(m)}\equiv-(2/\sqrt{-g})\times \delta \mathcal{S}_m/\delta
 g^{\mu\nu}$.
 In spatially flat FRW background Eq.(\ref{E1}) gives rise to the
following Friedmann equation

\beq
3H^2&=&\rho_m-c_1 \pi+\frac{c_2}{2}\dot{\pi}^2-3c_3 H\dot{\pi}^3\\
2\dot{H}+3H^2&=&-c_1 \pi-\frac{c_2}{2}\dot{\pi}^2-c_3\dot{\pi}^2\ddot{\pi}\\
\beta\rho_m&=&c_1-c_2\left(3H\dot{\pi}+\ddot{\pi}\right)+3c_3\dot{\pi}\left(3H^2\dot{\pi}+
\dot{H}\dot{\pi}+2H\ddot{\pi}\right) \label{F1} \eeq

It is interesting to note that Eq.(\ref{F1}) exhibits a self
accelerating solution given by

\beq
3H^2&=&-c_1 \pi+\frac{c_2}{2}\dot{\pi}^2-3c_3 H\dot{\pi}^3\\
&=&-c_1 \pi-\frac{c_2}{2}\dot{\pi}^2
\eeq

which means that $c_1=0$ (we assume $\dot{\pi}\neq 0$) and
$H^4=-c_2^3/54c_3^2$.
This last condition is impossible to satisfy as $c_2$ should be positive for stability of the theory.\\

We therefore conclude that a stable self accelerating solution, in
general, does not exist in the third order galileon gravity with
$(\nabla\pi)^2 \Box\pi$ term in the Einstein frame. It is therefore
necessary to invoke the higher order terms $L_4$ and $L_5$. In the
discussion, to follow, we shall demonstrate that the desired
solution can be obtained by adding the fourth order term in the
action (\ref{action1}). The analysis becomes cumbersome in the
presence of 5th order term which completes the Lagrangian of
galileon gravity. We have included the corresponding discussion and
results in the Appendix.

\section{Generalization to next higher order}

Let us consider the full covariant action of galileon gravity
\cite{Nicolis:2008in,Deffayet:2009wt}.

\be \mathcal{S}=\int{\rd^4
x}\sqrt{-g}\left(\frac{R}{2}+c_i~L^{(i)}\right)+\mathcal{S}_m[\psi_m,e^{2\beta\pi}g_{\mu\nu}]
\label{action2}
 \ee

where $\{c_i\}$ are constants and the $L_i^{'s}$ are given by

\beq
L^{(1)}&=&\pi\\
L^{(2)}&=&-\frac{1}{2}(\nabla\pi)^2 \equiv -\frac{1}{2} \pi_{;\mu}\pi^{;\mu}\\
L^{(3)}&=&-\frac{1}{2} (\nabla\pi)^2 \Box\pi\\
L^{(4)}&=&-\frac{1}{2}
(\nabla\pi)^2\left[(\Box\pi)^2-\pi_{;\mu\nu}\pi^{;\mu\nu}+\pi^{;\mu}\pi^{;\mu}G_{\mu\nu}\right]+(\Box\pi)\pi_{;\mu}\pi_{;\nu}\pi^{;\mu\nu}-\pi_{;\mu}\pi^{;\mu\nu}\pi_{;\nu\rho}\pi^{;\rho}
\eeq

Varying the action (\ref{action2}) with respect $\pi$ and the metric
$g_{\mu\nu}$, we obtained the field equation for $\pi$ and Einstein
equations\footnote{We have $T^{(4)}_{\mu\nu}=-T^{'(4)}_{\mu\nu}$ and
${\cal E}^{(4)}=-\frac{1}{2}{\cal E}^{'(4)}$ compared to
\cite{Deffayet:2009wt}.}

\beq
\label{eq:master1}
c_i{\cal E}^{(i)}&=&-\beta T^{(m)}\\
\label{eq:master2}
G_{\mu\nu}&=&T_{\mu\nu}^{(m)}+c_iT^{(i)}_{\mu\nu}
\eeq

where ${\cal E}^{(i)}=(1/\sqrt{-g}) \times \frac{\delta \mathcal{S}^{(i)}}{\delta \pi}$ and
$T^{(i)}_{\mu\nu}=-(2/\sqrt{-g})\times \delta \mathcal{S}^{(i)}/\delta g^{\mu\nu}$
with $S^{(i)}\equiv \int{\rd^4 x}\sqrt{-g} L^{(i)}$

where ${\cal E}^{'s}$ and $T^{(1) 's}_{\mu\nu}$ have the following form
\beq
{\cal E}^{(1)} &=& 1\\
{\cal E}^{(2)} &=& \Box\pi\\
{\cal E}^{(3)} &=& (\Box\pi)^2-\pi_{;\mu\nu}\pi^{;\mu\nu}-R^{\mu\nu}\pi_{;\mu}\pi_{;\nu}\\
{\cal E}^{(4)} &=& 2 \left(\Box \pi \right)^3
+4 \left(\pi_{;\mu}^{\hphantom{;\mu}\nu}\,\pi_{;\nu}^{\hphantom{;\nu}\rho}\,\pi_{;\rho}^{\hphantom{;\rho}\mu}\right)
-6 \left(\Box \pi\right) \left(\pi_{;\mu\nu}\pi^{;\mu\nu} \right)
- \left(\Box \pi\right) \left(\pi_{;\mu}\,\pi^{;\mu}\right) R
- 2 \left(\pi_{;\mu}\,\pi^{;\mu\nu}\,\pi_{;\nu}\right) R\nonumber\\
&&-4 \left(\Box \pi\right) \left(\pi_{;\mu} \,R^{\mu \nu}\,\pi_{;\nu}\right)
+ 2(\nabla\pi)^2\left(\pi_{;\mu\nu}\,R^{\mu \nu}\right)
+8 \left(\pi_{;\mu}\,\pi^{;\mu\nu}\,R_{\nu\rho}\,\pi^{;\rho}\right)
+ 4 \left(\pi_{;\mu}\,\pi_{;\nu}\,\pi_{;\rho\sigma}\,R^{\mu \rho \nu \sigma}\right)
\eeq

\beq
T^{(1)}_{\mu\nu}&=&\pi g_{\mu\nu}\\
T^{(2)}_{\mu\nu}&=&\pi_{;\mu}\pi_{;\nu}-\frac{1}{2}g_{\mu\nu}(\nabla\pi)^2\\
T^{(3)}_{\mu\nu}&=&\pi_{,\mu}\pi_{;\nu}\Box\pi+g_{\mu\nu}\pi_{;\lambda}\pi^{;\lambda\rho}\pi_{;\rho}-\pi^{;\rho}\left[\pi_{;\mu}\pi_{;\nu\rho}+\pi_{;\nu}\pi_{;\mu\rho}\right]\\
T^{(4)}_{\mu\nu}&=&
-4\left(\Box \pi\right)\pi^{;\rho}\bigl[\pi_{;\mu}\,\pi_{;\rho\nu}
+\pi_{;\nu}\,\pi_{;\rho\mu}\bigr]
+ 2 \left(\Box \pi\right)^2 \left(\pi_{;\mu}\,\pi_{;\nu}\right)
- 2 \left(\Box \pi\right)(\nabla\pi)^2\left(\pi_{;\mu\nu}\right)
- 4\left(\pi_{;\lambda}\,\pi^{;\lambda\rho}\,\pi_{;\rho}\right)\left(\pi_{;\mu\nu}\right)
\nonumber \\
&&
+ 4 \left(\pi^{;\lambda}\,\pi_{;\lambda\mu}\right)\left(\pi^{;\rho}\,\pi_{;\rho\nu}\right)
- 2 \left(\pi_{;\lambda\rho}\,\pi^{;\lambda\rho}\right)\left(\pi_{;\mu}\,\pi_{;\nu}\right)
+ 2 (\nabla\pi)^2\left(\pi_{;\mu}^{\hphantom{;\mu};\rho}\,\pi_{;\rho\nu}\right)
+ 4\,\pi_{;\lambda}\,\pi^{;\lambda\rho}\bigl[\pi_{;\rho\mu}\,\pi_{;\nu}+\pi_{;\rho\nu}\,\pi_{;\mu}\bigr]
\nonumber \\
&&
+\left(\Box \pi\right)^2 (\nabla\pi)^2g_{\mu \nu}
+ 4 \left(\Box \pi\right)\left(\pi_{;\lambda}\,\pi^{;\lambda\rho}\,\pi_{;\rho}\right) g_{\mu \nu}
- 4\left(\pi_{;\lambda}\,\pi^{;\lambda\rho}\,\pi_{;\rho\sigma}\,\pi^{;\sigma}\right)g_{\mu \nu}
- (\nabla\pi)^2\left(\pi_{;\rho\sigma}\,\pi^{;\rho\sigma}\right) g_{\mu \nu}
\nonumber \\
&&
- (\nabla\pi)^2\left(\pi_{;\mu}\,\pi_{;\nu}\right) R
+ \frac{1}{4}(\nabla\pi)^4g_{\mu \nu} R
+ 2 (\nabla\pi)^2 \pi^{;\rho} \bigl[R_{\rho\mu}\,\pi_{;\nu}+R_{\rho\nu}\,\pi_{;\mu}\bigr]
- \frac{1}{2}(\nabla\pi)^4R_{\mu \nu}
\nonumber \\
&&
- 2 (\nabla\pi)^2\left(\pi_{;\rho}\,R^{\rho \sigma}\,\pi_{;\sigma}\right)g_{\mu \nu}
+2 (\nabla\pi)^2\left(\pi^{;\rho}\,\pi^{;\sigma}\,R_{\mu\rho\nu\sigma}\right)
\eeq

It may be instructive to define the effective energy density and
pressure for $\pi$ matter. Indeed, for each $(i)$,

\be
\nabla^\mu T_{\mu\nu}^{(i)}=\pi_{;\nu}~{\cal E}^{(i)}
\ee

which allows us to write  the equation of conservation

\be
\nabla^\mu T_{\mu\nu}^{(m)}=\beta T^{(m)}~\pi_{;\nu}
\ee

For each $(i)$, assuming the perfect fluid form, we can express the
field energy momentum tensor as,
$T_{\mu\nu}^{(i)}=\left(\rho^{(i)}+P^{(i)}\right)u_\mu
u_\nu+P^{(i)}g_{\mu\nu}$ with $u_\mu\equiv -\sigma
\frac{\pi_{;\mu}}{\sqrt{-\left(\nabla\pi\right)^2}}$ and
$\sigma=\text{sign}\left(\pi_{;0}\right)$. The corresponding
expressions for $\rho^i$ and $P^i$ have following form

\begin{center}
\begin{tabular}{l l}
$\rho^{(1)} = -\pi$ & $~~~P^{(1)} = \pi$ \\
$\rho^{(2)} = -\frac{1}{2}(\nabla\pi)^2$ & $~~~P^{(2)} = -\frac{1}{2}(\nabla\pi)^2$ \\
$\rho^{(3)} = \pi_{;\lambda} \pi^{;\lambda \rho} \pi_{;\rho} -(\nabla\pi)^2 \Box \pi$ &
$~~~P^{(3)} = \pi_{;\lambda} \pi^{;\lambda \rho} \pi_{;\rho} $\\
$\rho^{(4)} = 6\Box \pi \pi_{;\lambda}\pi^{;\lambda\rho}\pi_{;\rho}-3\left(\Box \pi\right)^2\left(\nabla\pi\right)^2+3\left(\nabla\pi\right)^2 $ &
$~~~P^{(4)} = \left(\Box\pi\right)^2\left(\nabla \pi\right)^2 +4\Box \pi_{;\lambda}\pi^{;\lambda\rho}\pi_{;\rho}-4\pi_{;\lambda}\pi^{;\lambda\rho}\pi_{;\rho\sigma}\pi^{;\sigma}$\\
$~~~~~~~~~~+\frac{3}{4}R\left(\nabla\pi\right)^4-\frac{3}{2}\left(\nabla\pi\right)^2\pi^{;\rho}R_{\rho\sigma}\pi^{;\sigma}$ &
$~~~~~~~~~~-\left(\nabla\pi\right)^2\pi_{;\rho\sigma}\pi^{;\rho\sigma}+\frac{1}{4}R\left(\nabla\pi\right)^4-2\left(\nabla\pi\right)^2\pi^{;\rho}R_{\rho\sigma}\pi^{;\sigma}$\\
$~~~~~~~~~~-2\pi^{;\rho}\pi^{;\sigma}\pi^{;\mu}\pi^{;\nu}R_{\mu\rho\nu\sigma}-6\pi_{;\lambda}\pi^{;\lambda\rho}\pi_{;\rho\sigma}\pi^{;\sigma}$ &
$~~~~~~~~~~$
\end{tabular}
\end{center}
In the following section, we shall analyze the background solution
of the fourth order theory.
\section{Background dynamics}
Assuming the spatially flat background, we obtain evolution
 equations
of the fourth order galileon cosmology,

 \beq
\label{eq:Fried1}
3H^2&=&\rho_m+\frac{c_2}{2}\dot{\pi}^2-3c_3 H\dot{\pi}^3+\frac{45}{2}c_4 H^2 \dot{\pi}^4\\
\label{eq:Fried2}
2\dot{H}+3H^2&=&-\frac{c_2}{2}\dot{\pi}^2-c_3\dot{\pi}^2\ddot{\pi}+\frac{3}{2}c_4\dot{\pi}^3\left(3H^2\dot{\pi}+2\dot{H}\dot{\pi}+8H\ddot{\pi}\right)\\
\label{eq:KG}
\beta\rho_m&=&-c_2\left(3H\dot{\pi}+\ddot{\pi}\right)+3c_3\dot{\pi}\left(3H^2\dot{\pi}+\dot{H}\dot{\pi}
+2H\ddot{\pi}\right)-18c_4H\dot{\pi}^2\left(3H^2\dot{\pi}+2\dot{H}\dot{\pi}+3H\ddot{\pi}\right),
\eeq

where we have assumed, $c_1=0$ as we do not want include the
cosmological constant explicitly. In this case, the conservation has
standard form in presence of coupling $\beta$

\be
\dot{\rho}_m+3H\rho_m=\beta\rho_m \dot{\pi}
\ee

We may also define the total energy density and pressure for the
scalar field $\pi$

\beq
\rho_\pi&=&\frac{c_2}{2}\dot{\pi}^2-3c_3 H\dot{\pi}^3+\frac{45}{2}c_4 H^2 \dot{\pi}^4\\
P_\pi&=&\frac{c_2}{2}\dot{\pi}^2+c_3\dot{\pi}^2\ddot{\pi}-\frac{3}{2}c_4\dot{\pi}^3\left(3H^2\dot{\pi}+2\dot{H}\dot{\pi}+8H\ddot{\pi}\right)
\eeq

which can be used to check for the total equation of state parameter
$w_{\pi}=P_{\pi}/\rho_{\pi}$. In the next section, we discuss the
self accelerating solution of galileon cosmology.
\section{Self accelerating solution}

A self acceleration solution is characterized by $\rho_m=0$ and $H\equiv H_0=C^{st}$.

In this case, using equation (\ref{eq:Fried1}), we find  that
$\dot{\pi}\equiv \dot{\pi}_0=C^{st}$ and

\beq
H_0\dot{\pi}^\pm_0&=&\frac{c_3\pm\sqrt{c_3^2-8c_2c_4}}{12c_4}\\
48H_0^2&=&(\dot{\pi}^\pm_0)^2 A_{\pm}
\eeq

With $A_\pm=\frac{c_3^2-12c_2c_4\pm c_3\sqrt{c_3^2-8c_2c_4}}{c_4}$.

The existence of the self accelerating solution then implies the
following conditions on constants $c_1,c_2,c_3$ and $c_4$,

\beq
\label{dS:existence}
c_3^2-8c_2c_4&>&0\\
A_+ >0 &\text{or}& A_->0 \eeq

It is not difficult to check the stability of the solution.
 Let us consider the perturbation of the two self-accelerating
solutions,
 \be H=H_0+\delta
H,~~~~~~~~~~~~~~~~~~~~~~\dot{\pi}=\dot{\pi}_0+\delta \dot{\pi} \ee

It can easily be checked that $\dot{\delta H}=-3 H_0 \delta H$,
which means that the self-accelerating solutions are stable.

\section{Spherically symmetric solution}
We shall now be interested in  the spherically symmetric static
solution . We  consider a static point-like source of mass $M$,
located at the origin: $T^{(m)}=-M\delta^3(\overrightarrow{x})$ and
look for a spherically symmetric static solution for the field
$\pi(r)$ described by the following differential equation

\be \frac{c_2}{r^2}\frac{\rd}{\rd r}\left[r^2\pi
'(r)\right]+2\frac{c_3} {r^2}\frac{\rd}{\rd
r}\left[r\pi'(r)^2\right]+4\frac{c_4}{r^2}\frac{\rd}{\rd r}
\left[\pi'(r)^3\right]=\beta M\delta^3(\overrightarrow{x})
\label{Difure}
 \ee

Integration of Eq.(\ref{Difure}) gives the following relation,

\be
c_2\left(\frac{\pi'(r)}{r}\right)+2c_3\left(\frac{\pi'(r)}{r}\right)^2+
4c_4\left(\frac{\pi'(r)}{r}\right)^3=\beta\frac{r_s}{r^3}
\ee

Where $r_s$ is the Schwarzschild radius of the source.\\

The conditions of existence of the solution are derived following
Ref.\cite{Nicolis:2008in}:

if $\beta>0$ $\Rightarrow$  sign($c_2$)=sign($c_4$) and $c_3>-\sqrt{3c_2 c_4}$ which means that $c_3>\sqrt{8c_2 c_4}$ if we consider the condition (\ref{dS:existence})\\

if $\beta<0$ $\Rightarrow$  sign($c_2$)=sign($c_4$) and $c_3<\sqrt{3c_2 c_4}$ which means that $c_3<-\sqrt{8c_2 c_4}$ if we consider the condition (\ref{dS:existence})\\\\

%

In case $\beta<0$, at short distances, the solution is not analytic in the neighborhood of $r=0$ and
we shall not consider this case any further.

Whereas for $\beta>0$,

\be
\pi'(r)=\frac{\left(c_4^2 r_s \beta\right)^{1/3}}{2^{2/3}c_4}
\ee

Then the galileon-mediated force is suppressed compared to the gravitational force:

\be
\frac{F_\pi}{F_{\text{grav}}}=\left(\frac{r}{r_\star}\right)^{2} \ll 1, \qquad \text{with}~~~r_\star^3=\left(\frac{|c_4|}{2\beta}\right)^{1/2}r_s
\ee

At large distances, we have

\be
\frac{F_\pi}{F_{\text{grav}}}=2 \frac{\beta}{c_2}
\ee

If $\beta\simeq c_2$, the galileon field can lead to late time acceleration of universe.\\

\section{Stability}

In order to study the stability of the aforesaid static solutions,
we
 perturb the scalar field $\pi$: $\pi \rightarrow \pi+\phi$ in a
fixed metric $g_{\mu\nu}$. We have neglected the perturbations of
the metric induced by the perturbations of the scalar field $\phi$;
the method is referred to test field approximation.

In order to proceed with the test field approximation, let us
rewrite the quadratic term in $\phi$ in the action

\be
\label{eq:actionq}
\mathcal{S}_\phi=\int \sqrt{-g} {\rm d}^4 x~ c^i Z^{\mu \nu}_{(i)}\phi_{;\mu}\phi_{;\nu}
\ee

with

\beq
Z^{\mu \nu}_{(1)}&=&0\\
Z^{\mu \nu}_{(2)}&=&-\frac{1}{2}g^{\mu\nu}\\
Z^{\mu \nu}_{(3)}&=&\pi^{;\mu\nu}-g^{\mu\nu}\Box \pi\\
Z^{\mu \nu}_{(4)}&=&-2\pi^{;\mu} R^{\nu \rho}\pi_{;\rho}-2\pi^{;\nu} R^{\mu \rho}\pi_{;\rho}
-R^{\mu \nu}(\nabla\pi)^2+R\pi^{;\mu}\pi^{;\nu}+6\Box\pi \pi^{;\mu \nu}-6\pi^{;\mu\rho}\pi_{;\rho}^{~\nu}+2R^{\mu\rho\sigma\nu}\pi_{;\rho}\pi_{;\sigma}\\
&~&+g^{\mu\nu}\left(3\pi_{;\rho\sigma}\pi^{;\rho\sigma}-3(\Box \pi)^2+2R_{\rho \sigma}\pi^{;\rho}\pi^{;\sigma}+\frac{1}{2}R(\nabla\pi)^2\right)
\eeq

The equation of motion for perturbations that follow from action
(\ref{eq:actionq}) is

\be \label{eq:hyperbolic}
-2c^{i}Z^{\mu\nu}_{(i)}\phi_{;\mu\nu}-2c^{i}Z^{\mu\nu}_{~~;\mu}\phi_{;\nu}+8\beta^2\phi
T^{(m)}=0, \ee which we shall use in the subsequent sections.

\subsection{Cauchy-problem}

Following the theorem due to Leray \cite{Wald:1984rg}, the scalar field $\phi$
propagates causally in the effective metric $G^{\mu\nu}_{\rm eff}=-2c^{i}Z^{\mu\nu}_{(i)}$
if spacetime $(\mathcal{M},G^{\mu\nu}_{\rm eff})$ is globally hyperbolic.
A necessary condition but not sufficient is the requirement of the hyperbolicity of
the equation (\ref{eq:hyperbolic}) that is a Lorentzian signature of the effective metric $G^{\mu\nu}_{\rm eff}$.\\

For the static spherical solution, the hyperbolicity is defined by

\beq
c_2+2 c_3(2\pi'/r + \pi'') + 12 c_4 (\pi'/r + 2 \pi'')\pi'/r>0\\
c_2+4c_3\pi'/r+12 c_4 \left(\pi'/r\right)^2>0\\
c_2+2c_3(\pi'/r+\pi'')+12c_4\pi''\pi'/r>0
\eeq

At large distances, we obtain the following conditions

\beq
c_2-36\beta^2\frac{c_4}{c_2^2}\frac{r_s^2}{r^6}>0\\
c_2+4\beta\frac{c_3}{c_2}\frac{r_s}{r^3}>0\\
c_2-2\beta\frac{c_3}{c_2}\frac{r_s}{r^3}>0,
 \eeq

which  reduce to $c_2>0$ at very large scales.

At small distances, we need to impose the conditions, $c_4>0$ and
$c_3>0$.

For the de Sitter phase, the hyperbolicity is defined by

\beq
G^{0 0}_{\rm eff}&=&-\frac{1}{4}(A_\pm+4c_2)<0,\\
a^2 G^{1 1}_{\rm eff}&=&\frac{1}{36}(A_\pm-4c_2)>0,
\eeq

which implies that $A_\pm>4 c_2$.

 We should however emphasize that this solution is derived when the scalar field is dominant (de Sitter phase), therefore any small perturbation of the scalar field leads to a
perturbation of the metric and the test field approximation is then no longer true.

\subsection{Hamiltonian approach}
An alternative way to study the stability is related to the positive
definiteness of  Hamiltonian of the underlying theory.
In a locally inertial frame, the Hamiltonian is

\be
\mathcal{H}=-\frac{1}{2}G^{00}_{\rm eff}\dot{\phi}^2+\frac{1}{2}G^{kl}_{\rm eff}\phi_{,k}\phi_{,l}
\ee

The condition of hyperbolicity of equation (\ref{eq:hyperbolic}) is
sufficient for the Hamiltonian to be bounded from below. The
condition of hyperbolicity imposes an important restriction on sound
speed which we consider next.

\subsection{Speed of sound}

From the equation (\ref{eq:hyperbolic}), it is obvious to define the
"sound speed" $c_s^2$; the condition of hyperbolicity of the
equation restricts $c_s$ to real values $c_s^2>0$. It is
straightforward to see that the condition of
  $c_s$ to be real, restrict the signature of the effective metric to $(-,+,+,+)$ or
  $(+,-,-,-)$.

  However, if we also impose the
positivity of the Hamiltonian, we have to
 consider the effective metric with the same signature that as that of the original metric
 $g_{\mu\nu}$ which is $(-,+,+,+)$, in our case
 .
 This condition for non superluminal behavior of the scalar
field $\phi$ is expressed by $c_s^2<1$.

In case of the de Sitter phase, it is trivial to see that $c_s^2=\frac{A_\pm-4c_2}{9(A_\pm +4c_2)}<1$ (because of the conditions of stability of the theory ($c_2>0$ and $A_\pm>0$)). But the problem is more delicate for the spherically symmetric solution. Indeed, equation (\ref{eq:hyperbolic}) can be rewritten as

\be
G^{00}_{\rm eff}\ddot{\phi}+G^{11}_{\rm eff}\partial_r^2 \phi+G^{22}_{\rm eff}r^2\partial_\Omega^2 \phi + \text{first derivatives of}~  \phi + ...=0
\ee

where $\partial_\Omega^2$ is the angular part of the Laplacian.\\

Therefore we can define the speed of radial and angular excitations as follows,

\beq
c_r^2&=&-\frac{G^{11}_{\rm eff}}{G^{00}_{\rm eff}}=\frac{c_2+4 c_3 \pi'/r+12 c_4 \pi'^2/r^2}{c_2+2c_3(2\pi'/r+\pi'')+12 c_4(\pi'^2/r^2+2\pi'' \pi'/r)}\\
c_\Omega^2&=&-\frac{r^2 G^{22}_{\rm eff}}{G^{00}_{\rm eff}}=\frac{c_2+2c_3(\pi'/r+\pi'')+12c_4\pi''
\pi'/r}{c_2+2c_3(2\pi'/r+\pi'')+12c_4(\pi'^2/r^2+2\pi'' \pi'/r)},
\eeq

which at large distances gives rise to

\beq
c_r^2&\approx&1+4\beta\frac{c_3}{c_2^2}\frac{r_s}{r^3}\\
c_\Omega^2&\approx&1-2\beta\frac{c_3}{c_2^2}\frac{r_s}{r^3}, \eeq

whereas for small distances, we find

\beq
c_r^2&=&1\\
c_\Omega^2&\approx&\frac{c_3}{6c_4}\frac{r}{\pi'}
\eeq

It is clear that at large distances, we have a superluminal behavior
($c_r^2>1$) of the scalar field $\phi$ for the static spherically
solution, but this behavior is physically possible if the theory
does not have Closed Causal Curves (CCCs) which leads to paradoxes
\cite{Wald:1984rg,Babichev:2007dw}. It is known that if a spacetime
is stably causal, it does not possesses CCCs which means that a
global time can be defined. This is the case if we can define a
global time for the two metrics $g_{\mu\nu}$ and $G_{\mu\nu}$.

For the static spherically symmetric solution, we will consider the
Minkowsky time $\eta^{\mu\nu}\nabla_\mu t\nabla_\nu t=-1$.

Then

\be G^{\mu\nu}_{eff}\nabla_\mu t\nabla_\nu t=-c_2 -2 c_3(2\pi'/r +
\pi'') - 12 c_4 (\pi'/r + 2 \pi'')\pi'/r \label{G} \ee

Eq.(\ref{G}), at large distances,  reduces to

\be
G^{\mu\nu}_{eff}\nabla_\mu t\nabla_\nu t=-c_2+36\beta^2\frac{c_4}{c_2^2}\frac{r_s^2}{r^6}
\ee

which is negative iff $r^6>36\beta^2 r_s^2 c_4/c_2^3$.

If this condition is satisfied then the space time
$(M,G_{\mu\nu}^{\rm eff})$ is stably causal which means that no
closed timelike curves exist. We should emphasize that this
condition is satisfied if the equation (\ref{eq:hyperbolic}) is
hyperbolic.

\section{Metric perturbations}
Let us consider the perturbed FLRW spacetime with  scalar metric
perturbations in the longitudinal gauge

\be {\rd}s^2=-(1+2\phi){\rd t^2}+a^2\left(1-2\psi\right){\rd x}^2
\ee

The linear matter perturbations $\delta_m$ on super horizon scales
satisfy the evolution equation similar to the one in Einstein
gravity

\be
\ddot{\delta}_m+2H\dot{\delta}_m-\frac{G_{eff}}{2}\rho_m\delta_m=0
\ee

with the modified Newtonian constant,

\be
G_{eff}=1+\frac{2 \left(c_3 \dot{\pi }^2 +2
\beta \right)^2+c_4 N_4}{4 c_2-2 c_3^2 \dot{\pi }^4-16 c_3 H
\dot{\pi }-8 c_3 \ddot{\pi}+c_4 D_4}
\ee

where $N_4$ and $D_4$ are given by

 \beq N_4&=&14 c_2 \dot{\pi }^4+c_3^2 \dot{\pi }^8-88 c_3 H
\dot{\pi }^5+4 c_3 \beta \dot{\pi }^6+20 c_3 \dot{\pi
   }^4 \ddot{\pi}-64 H       \beta  \dot{\pi }^3-24 \beta ^2 \dot{\pi }^4+96 \beta  \dot{\pi }^2 \ddot{\pi}\nonumber\\
&+&c_4 \left(-9 c_2 \dot{\pi }^8-12
   c_3 H \dot{\pi }^9-54 c_3 \dot{\pi }^8 \ddot{\pi}+492 H^2 \dot{\pi }^6-96 H \beta  \dot{\pi }^7-48 H \dot{\pi }^5 \ddot{\pi}+168 \dot{H}
   \dot{\pi }^6+18 \beta ^2 \dot{\pi }^8-144 \beta  \dot{\pi }^6 \ddot{\pi}+288 \dot{\pi }^4 \ddot{\pi}^2\right)\nonumber\\
&+&c_4^2 \left(18 \left(11
   H^2-6 \dot{H}\right) \dot{\pi }^{10}+648 H \dot{\pi }^9 \ddot{\pi}\right)\\
D_4&=&-12 c_2 \dot{\pi }^4-c_3^2 \dot{\pi }^8+80 c_3 H \dot{\pi }^5-24 c_3 \dot{\pi }^4 \ddot{\pi}+8 \left(13 H^2+6
   \dot{H}\right) \dot{\pi }^2+96 H \dot{\pi } \ddot{\pi}\nonumber\\
&+&c_4 \left(9 c_2 \dot{\pi }^8+12 c_3 H \dot{\pi }^9+54 c_3
   \dot{\pi }^8 \ddot{\pi}-24 \left(17 H^2+6 \dot{H}\right) \dot{\pi }^6+288 H \dot{\pi }^5 \ddot{\pi}\right)\nonumber\\
&+&c_4^2 \left(18 \left(6 \dot{H}-11
   H^2\right) \dot{\pi }^{10}-648 H \dot{\pi }^9 \ddot{\pi}\right)
\eeq

The study of generic models of modified gravity shows that there is
a characteristic signature in the growth function $f=\frac{\rm d \ln
\delta_m}{\rm d \ln a}$ which can allow us to distinguish these
models from $\Lambda CDM$ and other dynamical dark energy models
within the frame work of Einstein gravity. We expect similar
features in galileon gravity. We shall address this important issue
in our future work.

\section{Conclusion}
In this paper, we have investigated galileon gravity in its general
form. The model consists of an effective field $\pi$ Lagrangian
consisting of five terms $\sum_1^5 {c_i L^i}$ added to
Einstein-Hilbert action such that the field equation are of second
order. In spatially flat FRW background, we set up the evolutions
equations in the model and examine the existence and stability of
self accelerating solutions. We point out that these solutions, in
general ($\dot{\pi}\ne 0$), are not stable in the third order
galileon theory. We extend the analysis to the fourth and fifth
order theory. In fourth order theory, self accelerating solutions
exist provided that $c_3^2-8c_2c_4>0$ and $A_+>0$ or $A_->0$. We
show that there is at least one stable self-accelerating solution in
this case. The analysis is cumbersome in case of 5th order theory
and we have included the corresponding results in the appendix. The
conclusions reached in fourth order galileon theory are shown to
hold in general. In case of the spherically symmetric static
solution, we find that the solution exists provided that
$c_3>\sqrt{8c_2c_4}$. The solution is stable and the fifth force can
lead the acceleration of the universe if we assume $\beta\simeq c_2$
and and $c_4>0$. We find as expected that the galileon force
mediated by the scalar field $\pi$ is negligibly small at small
scales, because of the non-linear terms in the Lagrangian. However,
the fifth force is of the order of the gravitational force at large
scales in case,
$\beta\simeq c_2$.\\
Subsequently, we investigated the stability issues associated with
the spherically symmetric solution. Using the fixed background
method, we found superluminal behavior of perturbations as was
noticed in \cite{Nicolis:2008in}. It is really interesting  that
despite the superluminal behavior, there exist static solutions
which  do not possess any Closed Causal Curve allowing to avoid
paradoxes related to micro-causality and making the solution
physically acceptable. The model has a well posed Cauchy problem and
no Closed Causal Curves exists in this model even if we have a
superluminal behavior of the perturbation of the scalar field in the
static spherically symmetric situation at large distances.

We have included brief discussion on the metric perturbations and
have set up the evolution equation for linear matter perturbation in
the galileon gravity. In our opinion, it is important to study the
growth function $f=\frac{\rm d \ln \delta_m}{\rm d \ln a}$ which can
provide a discriminating signature of galileon gravity; we defer
this analysis to our future work.
\section*{ACKNOWLEDGEMENTS}
We thank S. Deser for a useful comment. MS is supported by DST project No.SR/S2/HEP-002/2008.
R.~G. thanks CTP, Jamia Millia Islamia for hospitality where this work was carried out.

\begin{appendix}
\section{The full Lagrangian of galileon gravity: Extension of the model to the fifth order term, $L^{(5)}$}

We consider the term $L^{(5)}$ derived in \cite{Deffayet:2009wt}

\beq
L^{(5)}&=&-\frac{1}{2}(\nabla\pi)^2\left[\left(\Box \pi\right)^3
-3 \left(\Box \pi\right) \left(\pi_{;\mu\nu}\,\pi^{;\mu\nu}\right)
+2 \left(\pi_{;\mu}^{\hphantom{;\mu}\nu}\,\pi_{;\nu}^{\hphantom{;\nu}\rho}\,\pi_{;\rho}^{\hphantom{;\rho}\mu}\right)
-3 \left(\pi_{;\mu}\, \pi_{;\nu}\, \pi_{;\rho\sigma}\, R^{\mu\rho\nu\sigma}\right)\right.
\nonumber \\
&&\left. -18 \left(\pi_{;\nu}\,\pi^{;\nu\rho}\,R_{\rho\sigma}\,\pi^{;\sigma}\right)
+3 \left(\Box \pi\right)\left(\pi_{;\nu} \,R^{\nu \rho}\,\pi_{;\rho}\right)
+\frac{15}{2}(\nabla\pi)^2\left(\pi_{;\nu} \,\pi^{;\nu\rho}\,\pi_{;\rho}\right) R\right]
\nonumber \\
&&
+3 \left[\pi_{;\mu}\,\pi^{;\mu\nu}\,\pi_{;\nu\rho}\,\pi^{;\rho\lambda}\,\pi_{;\lambda}
-\left(\Box \pi\right)\left(\pi_{;\mu}\pi^{;\mu\nu}\,\pi_{;\nu\rho}\,\pi^{;\rho}\right)\right]
\nonumber \\
&&+\frac{3}{2}\left[ \left(\Box \pi\right)^2\left(\pi_{;\mu}\,\pi^{;\mu\nu}\,\pi_{;\nu}\right)
-\left(\pi_{;\mu\nu}\,\pi^{;\mu\nu}\right)\left(\pi_{;\rho}\,\pi^{;\rho\lambda}\,\pi_{;\lambda}\right)\right]
\eeq

then the equations (\ref{eq:master1},\ref{eq:master2}) are modified by

\beq
{\cal E}^{(5)}&=&
\frac{5}{2} \left(\Box\pi\right)^4
-15\left(\Box\pi\right)^2 \left(\pi_{;\mu \nu}\,\pi^{;\mu \nu}\right)
-\frac{15}{4} \left(\Box\pi\right)^2 (\nabla\pi)^2 R
-\frac{15}{2} \left(\Box\pi\right)^2 \left(\pi_{;\mu}\,R^{\mu \nu}\,\pi_{;\nu}\right)
\nonumber \\
&&+20\left(\Box\pi\right)\left(\pi_{;\mu}^{\hphantom{;\mu}\nu}\,\pi_{;\nu}^{\hphantom{;\nu}\rho}\,\pi_{;\rho}^{\hphantom{;\rho}\mu}\right)
-\frac{15}{2} \left(\Box\pi\right) \left(\pi_{;\mu}\,\pi^{;\mu \nu}\,\pi_{;\nu}\right) R
+15\left(\Box\pi\right) (\nabla\pi)^2\left(\pi_{;\nu \rho}\,R^{\nu \rho}\right)
\nonumber \\
&&+30 \left(\Box\pi\right)\left(\pi_{;\mu}\,\pi^{;\mu\nu}\,R_{\nu \rho}\,\pi^{;\rho}\right)
+15 \left(\Box\pi\right) \left(\pi_{;\mu}\,\pi_{;\nu}\,\pi_{;\rho \sigma}\,R^{\mu \rho \nu \sigma}\right)
+\frac{15}{2} \left(\pi_{;\mu \nu}\,\pi^{;\mu \nu}\right)^2
-15 \left(\pi_{;\mu \nu}\,\pi^{;\nu \rho}\,\pi_{;\rho \sigma}\,\pi^{;\sigma\mu}\right)
\nonumber \\
&&+\frac{15}{4}(\nabla\pi)^2 \left(\pi_{;\nu \rho}\,\pi^{;\nu \rho}\right) R
+\frac{15}{2} \left(\pi_{;\mu}\,\pi^{;\mu\nu}\,\pi_{;\nu\rho}\,\pi^{;\rho}\right) R
+\frac{15}{2} \left(\pi_{;\mu \nu}\,\pi^{;\mu \nu}\right)\left(\pi_{;\rho}\,R^{\rho \sigma}\,\pi_{;\sigma}\right)
\nonumber \\
&&+15\left(\pi_{;\mu}\,\pi^{;\mu \nu}\,\pi_{;\nu}\right)\left(\pi_{;\rho \sigma}\,R^{\rho \sigma}\right)
-15 (\nabla\pi)^2\left(\pi_{;\nu}^{\hphantom{;\nu}\rho}\,R_\rho^{\hphantom{\rho}\sigma}\,\pi_{;\sigma}^{\hphantom{;\sigma}\nu}\right)
-30 \left(\pi_{;\mu}\,\pi^{;\mu \nu}\,\pi_{;\nu\rho}\,R^{\rho \sigma}\,\pi_{;\sigma}\right)
\nonumber \\
&&-15 \left(\pi_{;\mu}\,\pi^{;\mu\nu}\,R_{\nu\rho}\,\pi^{;\rho\sigma}\,\pi_{;\sigma}\right)
-\frac{15}{2} (\nabla\pi)^2\left(\pi_{;\nu \rho}\,\pi_{;\sigma \lambda}\,R^{\nu \sigma \rho \lambda}\right)
-15 \left(\pi_{;\mu}\,\pi_{;\nu}\,\pi_{;\rho \sigma}\,\pi^{;\sigma}_{\hphantom{;\sigma}\lambda}\,R^{\mu \rho \nu \lambda}\right)
\nonumber \\
&&+30 \left(\pi_{;\lambda}\,\pi^{;\lambda}_{\hphantom{;\lambda}\mu}\,\pi_{;\nu \rho}\,\pi_{;\sigma}\,R^{\mu\nu\rho\sigma}\right)
+\frac{15}{4}(\nabla\pi)^2\left(\pi_{;\nu}\,R^{\nu \rho}\,\pi_{;\rho}\right) R
-\frac{15}{2} (\nabla\pi)^2\left(\pi_{;\nu}\,R^{\nu\rho}\,R_{\rho \sigma}\,\pi^{;\sigma}\right)
\nonumber \\
&&-\frac{15}{2}(\nabla\pi)^2\left(\pi_{;\nu}\,\pi_{;\rho}\,R_{\sigma \lambda}\,R^{\nu \sigma \rho \lambda}\right)
+\frac{15}{4}(\nabla\pi)^2\left(\pi_{;\nu}\,\pi_{;\rho}\,R^\nu_{\hphantom{\nu}\sigma\kappa\lambda}\,R^{\rho \sigma \kappa \lambda}\right)
\eeq

\beq
{T}^{(5)}_{\mu\nu}&=&
\frac{5}{2} \left(\Box\pi\right)^3\left(\pi_{;\mu}\,\pi_{;\nu}\right)
+\frac{5}{2} \left(\Box\pi\right)^3(\nabla\pi)^2g_{\mu \nu}
-\frac{15}{2} \left(\Box\pi\right)^2(\nabla\pi)^2 \left(\pi_{;\mu \nu}\right)
-\frac{15}{2} \left(\Box\pi\right)^2\pi^{;\rho}\bigl[\pi_{;\rho\mu}\,\pi_{;\nu}+\pi_{;\rho\nu}\,\pi_{;\mu}\bigr]
\nonumber \\
&&+\frac{15}{2} \left(\Box\pi\right)^2\left(\pi_{;\rho}\,\pi^{;\rho \sigma}\,\pi_{;\sigma}\right)g_{\mu \nu}
+15 \left(\Box\pi\right)(\nabla\pi)^2\left(\pi_{;\mu\sigma}\,\pi^{;\sigma}_{\hphantom{;\sigma}\nu}\right)
-15 \left(\Box\pi\right)\left(\pi_{;\rho}\,\pi^{;\rho \sigma}\,\pi_{;\sigma}\right)\left(\pi_{;\mu \nu}\right)
\nonumber \\
&&-\frac{15}{2} \left(\Box\pi\right)\left(\pi_{;\rho \sigma}\,\pi^{;\rho \sigma}\right)\left(\pi_{;\mu}\,\pi_{;\nu}\right)
+15 \left(\Box\pi\right)\left(\pi^{;\rho}\,\pi_{;\rho\mu}\right)\left(\pi^{;\sigma}\,\pi_{;\sigma\nu}\right)
+15 \left(\Box\pi\right)\pi_{;\rho}\,\pi^{;\rho\sigma}\bigl[\pi_{;\sigma\mu}\,\pi_{;\nu}+\pi_{;\sigma\nu}\,\pi_{;\mu}\bigr]
\nonumber \\
&&-\frac{15}{2} \left(\Box\pi\right)(\nabla\pi)^2\left(\pi_{;\sigma \lambda}\,\pi^{;\sigma \lambda}\right)g_{\mu \nu}
-15 \left(\Box\pi\right)\left(\pi_{;\rho}\,\pi^{;\rho\sigma}\,\pi_{;\sigma \lambda}\,\pi^{;\lambda}\right)g_{\mu \nu}
-\frac{15}{4} \left(\Box\pi\right)(\nabla\pi)^2\left(\pi_{;\mu}\,\pi_{;\nu}\right) R
\nonumber \\
&&+\frac{15}{2} \left(\Box\pi\right)(\nabla\pi)^2\pi^{;\sigma}\bigl[R_{\sigma\mu}\,\pi_{;\nu}+R_{\sigma\nu}\,\pi_{;\mu}\bigr]
-\frac{15}{2} \left(\Box\pi\right)(\nabla\pi)^2\left(\pi_{;\sigma}\,R^{\sigma \lambda}\,\pi_{;\lambda}\right)g_{\mu \nu}
\nonumber \\
&&+\frac{15}{2} \left(\Box\pi\right)(\nabla\pi)^2\left(\pi^{;\sigma}\,\pi^{;\lambda}\,R_{\mu\sigma\nu\lambda}\right)
+\frac{15}{2}(\nabla\pi)^2\left(\pi_{;\sigma \lambda}\,\pi^{;\sigma \lambda}\right)\left(\pi_{;\mu \nu}\right)
-15 (\nabla\pi)^2\left(\pi_{;\mu\sigma}\,\pi^{;\sigma\lambda}\,\pi_{;\lambda\nu}\right)
\nonumber \\
&&+15 \left(\pi_{;\rho}\,\pi^{;\rho \sigma}\,\pi_{;\sigma}\right)\left(\pi_{;\mu\lambda}\,\pi^{;\lambda}_{\hphantom{;\lambda}\nu}\right)
+15 \left(\pi_{;\rho}\,\pi^{;\rho\sigma}\,\pi_{;\sigma \lambda}\,\pi^{;\lambda}\right)\left(\pi_{;\mu \nu}\right)
+5\left(\pi_{;\rho}^{\hphantom{;\rho}\sigma}\,\pi_{;\sigma}^{\hphantom{;\sigma}\lambda}\,\pi_{;\lambda}^{\hphantom{;\lambda}\rho}\right)\left(\pi_{;\mu}\,\pi_{;\nu}\right)
\nonumber \\
&&+\frac{15}{2}\left(\pi_{;\sigma \lambda}\,\pi^{;\sigma \lambda}\right)\pi^{;\rho}\bigl[\pi_{;\rho\mu}\,\pi_{;\nu}+\pi_{;\rho\nu}\,\pi_{;\mu}\bigr]
-15\,\pi^{;\rho}\,\pi_{;\rho \sigma}\,\pi^{;\sigma \lambda}\bigl[\pi_{;\lambda\mu}\,\pi_{;\nu}+\pi_{;\lambda\nu}\,\pi_{;\mu}\bigr]
\nonumber \\
&&-15\,\pi_{;\rho}\,\pi^{;\rho\lambda}\,\pi^{;\sigma}\bigl[\pi_{;\lambda\mu}\,\pi_{;\sigma\nu}+\pi_{;\lambda\nu}\,\pi_{;\sigma\mu}\bigr]
+5 (\nabla\pi)^2\left(\pi_{;\sigma}^{\hphantom{;\sigma}\lambda}\,\pi_{;\lambda}^{\hphantom{;\lambda}\kappa}\,\pi_{;\kappa}^{\hphantom{;\kappa}\sigma}\right)g_{\mu \nu}
-\frac{15}{2}\left(\pi_{;\rho}\,\pi^{;\rho \sigma}\,\pi_{;\sigma}\right)\left(\pi_{;\lambda \kappa}\,\pi^{;\lambda \kappa}\right)g_{\mu \nu}
\nonumber \\
&&+15 \left(\pi_{;\rho}\,\pi^{;\rho\sigma}\,\pi_{;\sigma\lambda}\,\pi^{;\lambda\kappa}\,\pi_{;\kappa}\right)g_{\mu \nu}
+\frac{15}{4}(\nabla\pi)^2\pi^{;\sigma}\bigl[\pi_{;\sigma\mu}\,\pi_{;\nu}+\pi_{;\sigma\nu}\,\pi_{;\mu}\bigr] R
-\frac{15}{4} (\nabla\pi)^2\left(\pi_{;\sigma}\,\pi^{;\sigma \lambda}\,\pi_{;\lambda}\right)R\,g_{\mu \nu}
\nonumber \\
&&+\frac{15}{2} (\nabla\pi)^2\left(\pi_{;\sigma}\,\pi^{;\sigma \lambda}\,\pi_{;\lambda}\right)R_{\mu \nu}
+\frac{15}{2} (\nabla\pi)^2\left(\pi_{;\sigma}\,R^{\sigma \lambda}\,\pi_{;\lambda}\right)\left(\pi_{;\mu \nu}\right)
+\frac{15}{2} (\nabla\pi)^2\left(\pi_{;\sigma\lambda}\,R^{\sigma\lambda}\right)\left(\pi_{;\mu}\,\pi_{;\nu}\right)
\nonumber \\
&&-\frac{15}{2} (\nabla\pi)^2\pi_{;\sigma}\,\pi^{;\sigma\lambda}\bigl[R_{\lambda\mu}\,\pi_{;\nu}+R_{\lambda\nu}\,\pi_{;\mu}\bigr]
-\frac{15}{2} (\nabla\pi)^2\pi^{;\lambda}\,\pi^{;\sigma}\bigl[R_{\lambda\mu}\,\pi_{;\sigma\nu}+R_{\lambda\nu}\,\pi_{;\sigma\mu}\bigr]
\nonumber \\
&&-\frac{15}{2} (\nabla\pi)^2\pi_{;\sigma}\,R^{\sigma \lambda}\bigl[\pi_{;\lambda\mu}\,\pi_{;\nu}+\pi_{;\lambda\nu}\,\pi_{;\mu}\bigr]
+15(\nabla\pi)^2\left(\pi_{;\sigma}\,\pi^{;\sigma\lambda}\,R_{\lambda\kappa}\,\pi^{;\kappa}\right)g_{\mu \nu}
\nonumber \\
&&-\frac{15}{2} (\nabla\pi)^2\pi^{;\sigma}\,\pi^{;\lambda \kappa}\bigl[R_{\mu\lambda\sigma\kappa}\,\pi_{;\nu}+R_{\nu\lambda\sigma\kappa}\,\pi_{;\mu}\bigr]
+\frac{15}{2} (\nabla\pi)^2\pi^{;\sigma}\,\pi^{;\lambda}\bigl[R_{\mu\sigma \lambda \kappa}\,\pi^{;\kappa}_{\hphantom{;\kappa}\nu}+ R_{\nu\sigma \lambda \kappa}\,\pi^{;\kappa}_{\hphantom{;\kappa}\mu}\bigr]
\nonumber \\
&&-\frac{15}{2} (\nabla\pi)^2\pi_{;\sigma}\,\pi^{;\sigma\lambda}\,\pi^{;\kappa}\bigl[R_{\mu\lambda\nu\kappa}+R_{\nu\lambda\mu\kappa}\bigr]
+\frac{15}{2} (\nabla\pi)^2\left(\pi_{;\sigma}\,\pi_{;\lambda}\,\pi_{;\kappa\tau}\,R^{\sigma\kappa\lambda\tau}\right)g_{\mu \nu}.
\eeq

The Friedmann equations for this model are\\

\beq
3H^2&=&\rho_m+\frac{c_2}{2}\dot{\pi}^2-3c_3 H\dot{\pi}^3+\frac{45}{2}c_4 H^2 \dot{\pi}^4-\frac{105}{2}c_5  H^3 \dot{\pi}^5\\
2\dot{H}+3H^2&=&-\frac{c_2}{2}\dot{\pi}^2-c_3\dot{\pi}^2\ddot{\pi}+\frac{3}{2}c_4\dot{\pi}^3\left(3H^2\dot{\pi}+2\dot{H}\dot{\pi}+8H\ddot{\pi}\right)-\frac{15}{2}c_5 H \dot{\pi}^4\left(2H^2\dot{\pi}+2\dot{H}\dot{\pi}+5H\ddot{\pi}\right)\\
\beta\rho_m&=&-c_2\left(3H\dot{\pi}+\ddot{\pi}\right)+3c_3\dot{\pi}\left(3H^2\dot{\pi}+\dot{H}\dot{\pi}+2H\ddot{\pi}\right)-18c_4H\dot{\pi}^2\left(3H^2\dot{\pi}+2\dot{H}\dot{\pi}+3H\ddot{\pi}\right) \nonumber \\
& &+\frac{75}{2}c_5H^2\dot{\pi}^3\left(3H^2\dot{\pi}+3\dot{H}\dot{\pi}+4H\ddot{\pi}\right)
\eeq

Therefore the self accelerating solution exist if there is a real solution of the equation

\beq
c_2-3c_3X+18c_4X^2-\frac{75}{2}c_5X^3&=&0\\
c_2-9c_4X^2+30c_5X^3&<&0,~~\text{with}~~~X=H_0\dot{\pi}_0
\eeq

If we have a solution of this system therefore the self accelerating solution is stable. In fact if we consider a perturbation of
the self accelerating solution $H=H_0+\delta H$ and $\dot{\pi}=\dot{\pi}_0+\delta \dot{\pi}$, it is straightforward to see that, $\dot{\delta H}=-3 H_0 \delta H$.

We found that the spherical symmetric solution is not modified by the fifth term as it was noticed in Ref.\cite{Nicolis:2008in}.
\end{appendix}

\end{document}